\documentclass[twocolumn,amsmath,amssymb,floatfix,superscriptaddress,aps,prl]{revtex4-2}

\usepackage{physics,comment}
\usepackage[colorlinks=true,linkcolor=blue,citecolor=blue]{hyperref}
\usepackage{bm,graphicx,epsfig}
\usepackage[usenames]{color}
\usepackage[english]{babel}
\usepackage{amsfonts,stmaryrd,amssymb,comment}
\usepackage{enumerate} 
\usepackage[ruled]{algorithm2e} 
\usepackage[framemethod=tikz]{mdframed} 
\usepackage{amsthm,amsmath,amssymb,amsfonts,graphicx,verbatim,xcolor,bm} 
\usepackage[normalem]{ulem}

\usepackage{graphicx,amsmath,comment} 
\usepackage[margin=1.0in]{geometry}
\usepackage{comment,ulem,amsmath,bbm}

\newcommand{\papertitle}{Control transition in a temporally random classical spin chain}

\begin{document}

\title{\papertitle}

\author{Elisha Shmalo}
\affiliation{Department of Physics and Astronomy, Center for Materials Theory, Rutgers University, Piscataway, New Jersey 08854, USA}

\author{J. H. Pixley}
\affiliation{Department of Physics and Astronomy, Center for Materials Theory, Rutgers University, Piscataway, New Jersey 08854, USA}
\affiliation{Center for Computational Quantum Physics, Flatiron Institute, New York, New York 10010, USA}

\author{Manas Kulkarni}
\affiliation{International Centre for Theoretical Sciences (ICTS-TIFR),
Tata Institute of Fundamental Research, Bangalore 560089, India}

\author{Sarang Gopalakrishnan}
\affiliation{Department of Electrical and Computer Engineering,
Princeton University, Princeton, NJ 08544, USA}

\author{David A. Huse}
\affiliation{Department of Physics, Princeton University, Princeton, New Jersey 08544, USA}

\begin{abstract}
We theoretically explore a phase transition between controlled and chaotic dynamics in a classical spin chain model subject to chaotic Hamiltonian dynamics and a contractive ``control map'', which alternate 
in time. The control map drives the system toward a target configuration that is an unstable fixed point under the chaotic dynamics.  When the control is strong enough, the target configuration is the globally attracting stable fixed point of the dynamics; for weaker control, the many-body dynamics remains chaotic for almost all initial states.  The phase transition between controlled and chaotic phases has a mixed character:  As the transition is approached from the chaotic phase, the fraction of the spins that are far from the target configuration goes continuously to zero, and there are  diverging spatial and temporal correlation lengths; however, the leading Lyapunov exponent is discontinuous across the transition, jumping from a positive value in the chaotic phase to a negative value in the controlled phase.  We present evidence that this transition is in the same universality class as directed percolation in the presence of temporal randomness, a universality class for which we 
obtain results that are consistent with the dynamical Harris criterion but do not saturate the bound.
\end{abstract}

\maketitle

Controlling chaotic dynamical~\cite{ott2002chaos, strogatz2001nonlinear,B2000,GD93,EB2003} systems through feedback is a central question in non-equilibrium dynamics due to its  far reaching applications in robotics~\cite{de2018mimicking}, computing~\cite{sinha1998dynamics}, and statistical mechanics~\cite{robledo2013generalized}. 
Despite the exponential sensitivity of chaotic dynamics to perturbations, 
adding strong enough dissipation  
can suppress chaos~\cite{PhysRevLett.64.1196,pyragas1992continuous,antoniou1997probabilistic}.  In some chaotic systems, it is possible to controllably maintain the system in a desired state by \emph{engineering} the dissipation to target that state~\cite{krechetnikov2007dissipation}. 
The case in which the target state is an unstable fixed point of the chaotic dynamics is especially simple: in this case, the fixed point can change from unstable to stable at the transition from chaos to stable control~\cite{PhysRevLett.64.1196}. Here, the transition out of chaos is a type of ``absorbing-state'' transition~\cite{tauber2014critical}: the fixed point itself is invariant under the dynamics for any parameters, but typical initial conditions only reach the fixed point in the controlled phase. There has also been recent progress in mitigating chaos via imposition of dynamical constraints~\cite{SR1,SR2} and stochastic resetting~\cite{AK1,AK2}.

Recently, absorbing-state and other control transitions have been revisited from the perspective of monitored and open-system quantum dynamics, leading to the discovery of control-induced phase transitions (CIPTs)~\cite{iadecola2023measurement,lemaire2024separate,pan2024local,piroli2023triviality, sierant2023controlling,odea2024entanglement,ravindranath2023entanglement,thompson2024population,chertkov2023characterizing,pokharel2025order,MP2026}, which arise from embedding  a classical dynamical phase transition into a quantum system.
The classical transitions themselves remain imperfectly understood for strongly-interacting systems. However, compared with the standard models of absorbing-state transitions (which are automata defined on discrete state spaces)~\cite{tauber2014critical}, Hamiltonian dynamics offers new diagnostic 
quantities, including the leading Lyapunov exponent~\cite{pikovsky2016lyapunov}, whose scaling behavior at absorbing-state transitions has not yet been considered.

\begin{figure}[t!]
\includegraphics[clip, trim=5 10 0 10, width=0.45\textwidth]{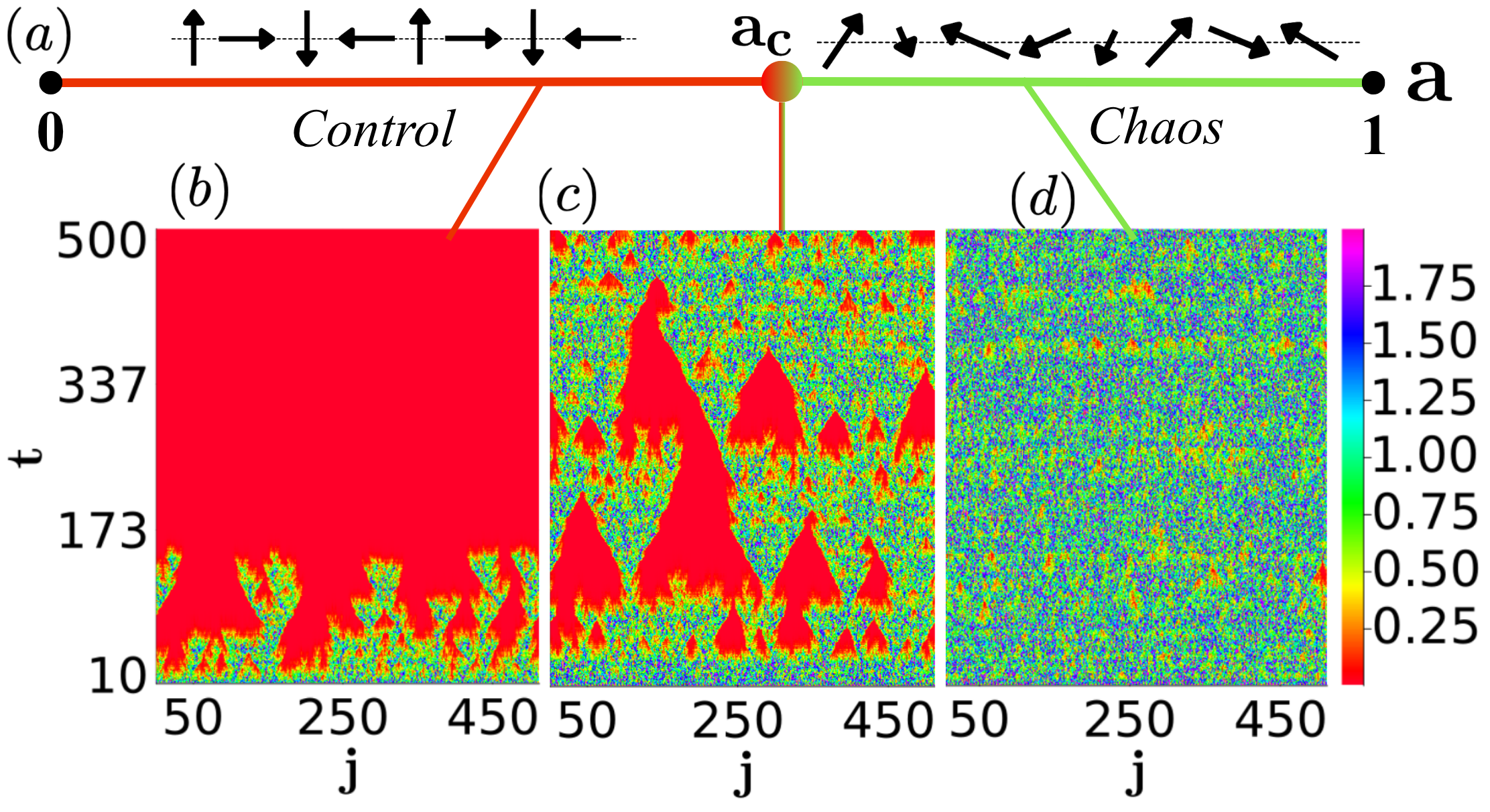}
\caption{Top Row: (a) Schematic of the phase diagram between controlled and chaotic phases tuned by the control parameter $a$, separated by the critical point $a_c$. The wavelength-4 spin spiral in Eq.~\eqref{eqn:spinspiral} is depicted, as well as a chaotic configuration. Bottom Row:
activity dynamics after starting in a random initial state, revealed through space-time ($j$-$t$) plots of $\delta \mathcal{S}_j(t)$ defined in Eq.~\eqref{eq:Sdiff}
(values given in color bar).  In the controlled phase (b), $a=0.72$, the system goes to the controlled state (red) at  
long times, note how the temporal randomness causes many active regions to ``die'' simultaneously; in the critical regime (c), $a=0.758\approx a_c$, we have ``active'' and ``dead'' regions and the  
long range spatial correlations remain apparent; 
deeper in the chaotic phase (d), $a=0.8$, it looks quite random.
}
\label{fig:model}
\end{figure}

In the present work, we explore the critical properties of the absorbing-state transition for a model in which classical Hamiltonian dynamics and a contractive control map alternate in time. The Hamiltonian dynamics is generated by a Heisenberg spin-chain Hamiltonian, while the control map drives the dynamics toward  
a spin spiral configuration (see Fig.~\ref{fig:model})
that is an unstable fixed point of the Hamiltonian dynamics. Compared with the ``standard'' absorbing-state transition (for example, in Stavskaya's automaton~\cite{stavskaya1968homogeneous,shnirman1968ergodicity,toom1968family}), this differs in two ways: first, the state space is continuous; and second, the dynamics is temporally random, with the signs of some couplings chosen randomly in each time step. 
This second modification is made in order to avoid persistent non-chaotic features, e.g. solitons~\cite{Tjon-1977,McRoberts-2022} from arising (see Supplemental Material). 

We study the phase transition between the  contractive (or ``controlled'') phase (where the spiral state is the global attractor) and the chaotic one (where almost all initial states remain chaotic~\cite{WijnLargestLyapExpClassicalSpin, DasDrivenHeisenberg}). We find that this transition has a mixed character: Correlation lengths in the chaotic phase diverge continuously with nontrivial exponents as one approaches the transition, and 
in the steady-state distribution the fraction of ``active'' spins that deviate strongly from the control state  
goes continuously to zero, suggesting a nonequilibrium critical point.  However, there are no critical fluctuations in the controlled phase, and the leading Lyapunov exponent jumps discontinuously at the transition: it limits to a value strictly greater (less) than zero as the transition is approached from the chaotic (controlled) side. We gain further insight into the nature of this transition by comparing the critical behavior with that of a standard model of directed percolation (Stavskaya's automaton~\cite{stavskaya1968homogeneous,shnirman1968ergodicity,toom1968family}) subject to temporally random variations of its parameters. The critical exponents we find are consistent in the two models, suggesting that both belong to a universality class that has until now only been loosely explored: temporally-random directed percolation~\cite{IwanDPtemprand}.

 \begin{figure*}[t!]
   \centering
\includegraphics[clip, trim=39 0 9 10, width=0.32\textwidth]{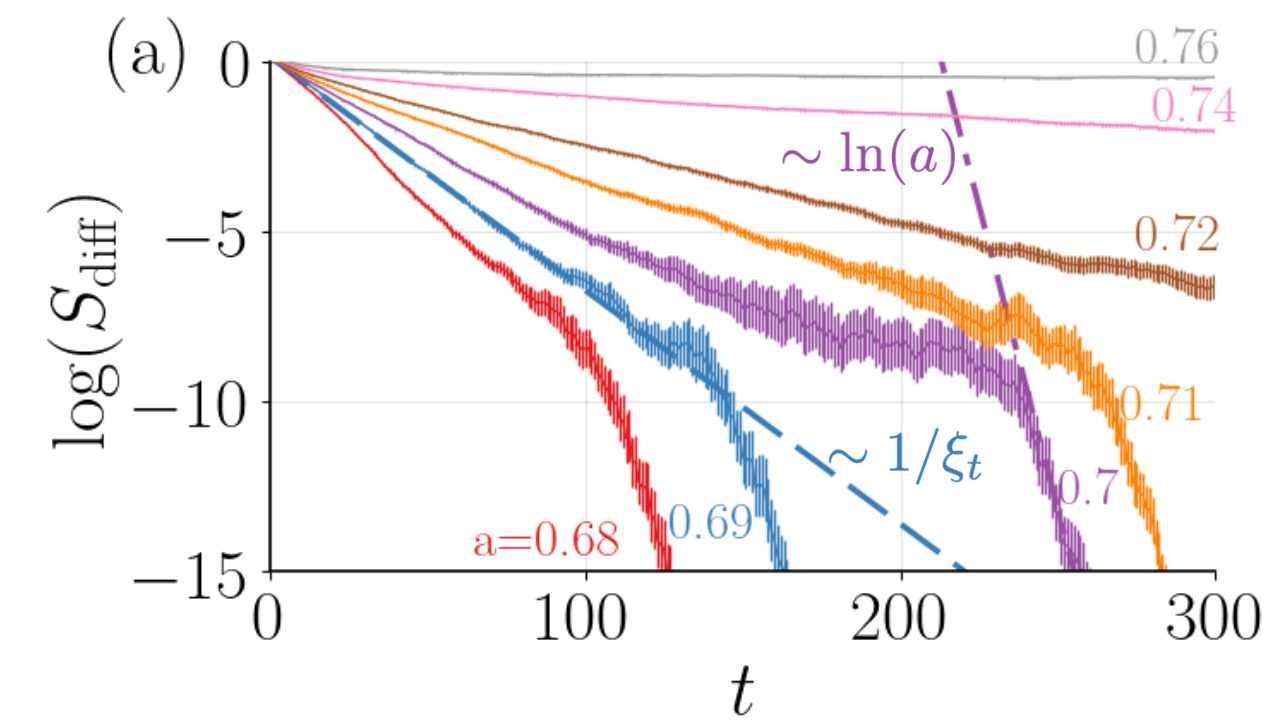}
\includegraphics[clip, trim=5 0 50 20, width=0.32\textwidth]{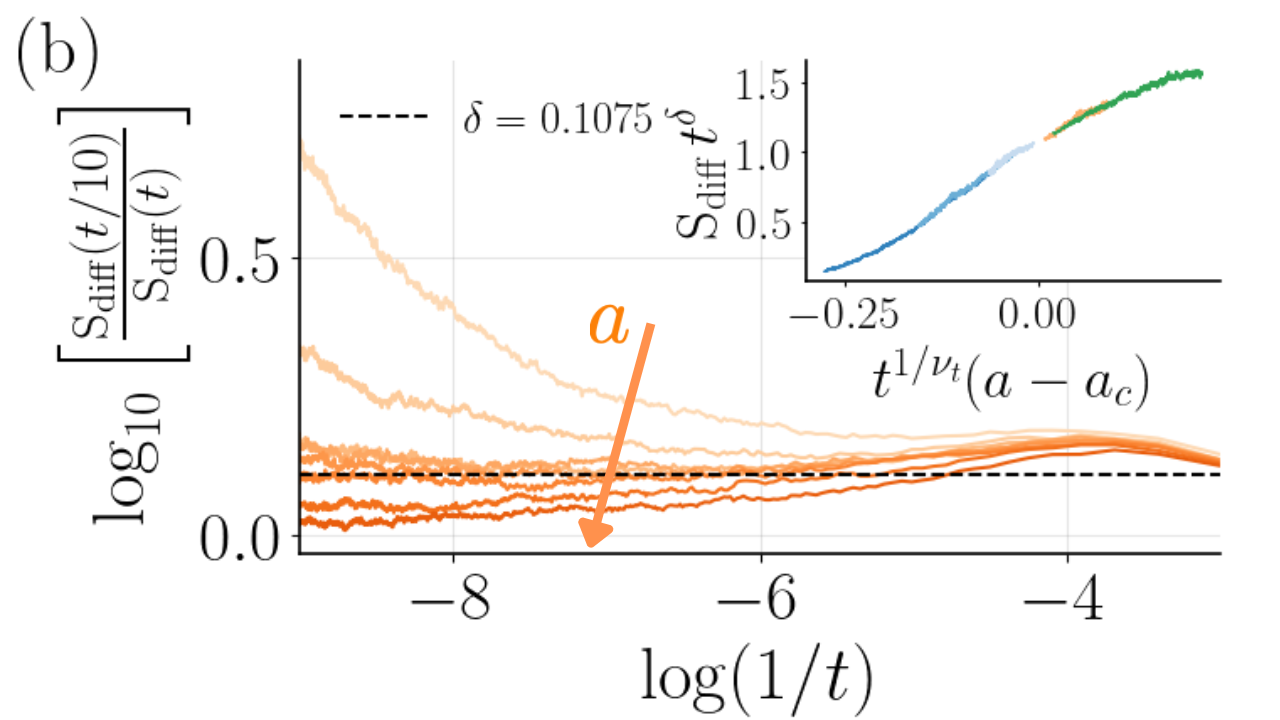}
\includegraphics[clip, trim=10 0 40 10, width=0.32\textwidth]{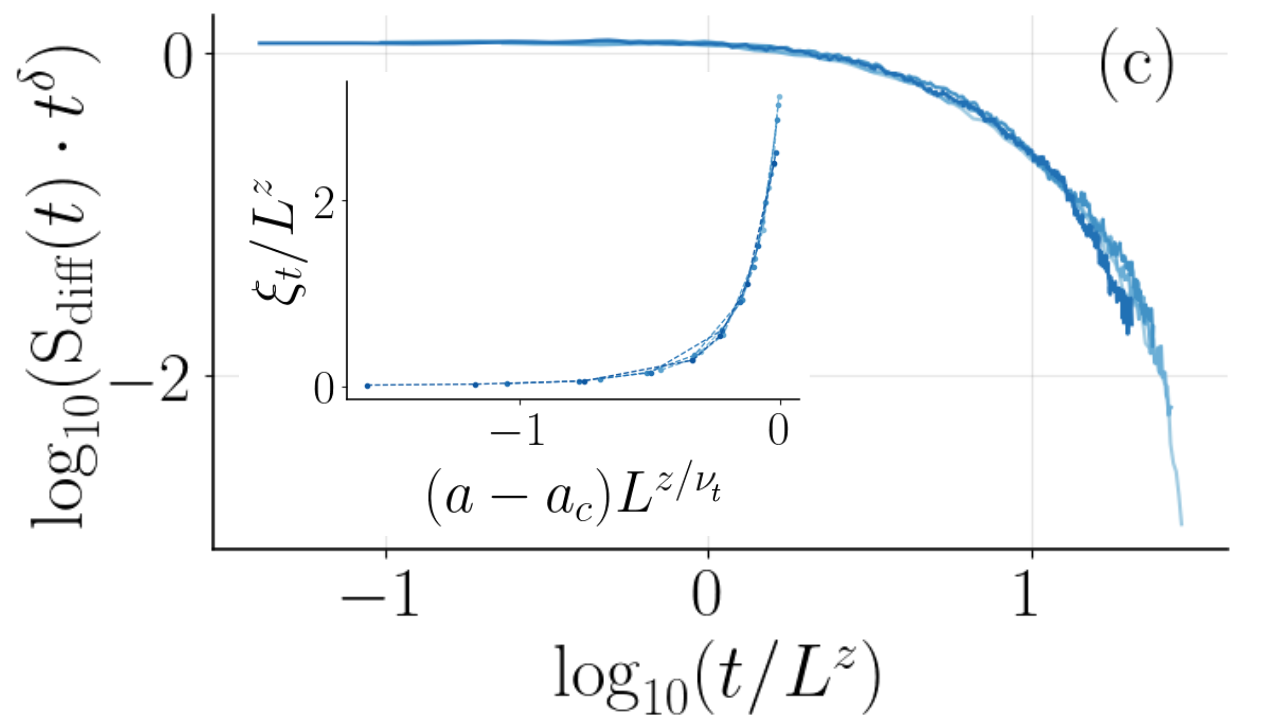}
\caption{ {\bf Activity across the transition}: (a) The activity $S_{\mathrm{diff}}(t)$ in Eq.~\eqref{eq:Sdiff} on a log-scale versus time ($t$) for chain length $L=256$  and $a=0.68$ to $a=0.76$ across the controlled and chaotic phases. In   the controlled phase two  temporal regimes are separated by $t^*$ [see Eq.~\eqref{eq:t*2}], displaying the critical decay regime set by $\xi_t$ (example in blue dashed line) and the later-time decay that occurs once all realizations are close enough to the control state  
(purple dotted-dashed line). (b) ~$\log_b\!\left[S_{\mathrm{diff}}(t/b)/S_{\mathrm{diff}}(t)\right]$ with $b = 10$, plotted as a function of time. The $a$ values range from $a=0.7535$ (highest curve) to $a=0.7615$ (lowest curve). Finding the value of $a$ for which this ratio is constant in time at late times determines $a_c = 0.7580(3)$ and $\delta = 0.11(2)$. Inset shows scaling collapse near $a_c = 0.7578$, with $a \in \{0.7535 (\text{dark blue}), 0.7555(\text{blue}), 0.757(\text{light blue}), 0.7595(\text{orange}), 0.7615(\text{green})\}$,  to determine $\nu_t = 2.2(2)$. 
(c) Finite-size scaling of the activity for $L \in \{64, 128, 256, 512\}$ using Eq.~\eqref{eq:general_scaling} at $a_c = 0.7582$ to determine $z \approx 1.3(2)$. The inset shows finite-size scaling of $\xi_t(a)$, focusing on times $t\ll t^*$, using the same $L$ values. Using Eq.~(\ref{eq:xi-scaling}) with $a_c = 0.7580$ and $\nu_t = 2.24$ (from Fig.~\ref{fig:Sdiff_vs_t}), yields $z\approx 1.3(2)$.}
\label{fig:Sdiff_vs_t}
\end{figure*}
{\it Model construction and observables}:  The dynamics of our model alternates between Hamiltonian time evolution and a dissipative control step:
The Hamiltonian part of the dynamics is produced by the spin chain Hamiltonian
\begin{equation}
    H = -\sum_{i=1}^L \left( J_x(t)S^x_iS^x_{i+1}+J_y(t)S^y_iS^y_{i+1}+S^z_iS^z_{i+1}\right)
    \label{eq:ham}
\end{equation}
with periodic boundary conditions (${\bf S}_{L+1}\equiv {\bf S}_1)$. 
Each spin is a classical O(3) vector that is normalized to unity $|{\bf S}_i|=1$.  We take $J_x$ and $J_y$ to independently and randomly take values $\pm 1$ during each time step's Hamiltonian evolution, in order to eliminate solitons that appear in the dynamics near the transition if they are instead taken nonrandom (see Supplemental Material).  Note that this time-dependent $H$ remains translationally-invariant at all times. 
Denoting $\tilde {\bf S}_j=(J_x S_j^x,J_y S_j^y,S_j^z)$ with $ {\bf S}_j=(S_j^x, S_j^y,S_j^z)$,
the time evolution of the spin at site $j$ is
\begin{eqnarray}
    \frac{d{\bf S}_j}{d\tau}&=\{{\bf S}_j,H\}=&-(\tilde {\bf S}_{j-1}+\tilde {\bf S}_{j+1})\times {\bf S}_{j}~,  
    \label{eq:EOM}
\end{eqnarray}
where $\{\cdot,\cdot\}$ denotes a Poisson bracket, and this dynamics acts for a time duration of one unit of time ($\tau=1/J)$ during the Hamiltonian portion of each time step.

To control the chaotic dynamics, we choose an 
unstable fixed point of the Hamiltonian dynamics to stabilize.  We use the spin spiral:  
\begin{equation}
    {\bf S}_j^{0}=\left(0,\cos(\pi j/2 ),\sin(\pi j/2 )\right)~,
    \label{eqn:spinspiral}
\end{equation}
with spatial period four sites, see Fig.~\ref{fig:model}.  This fixed point is marginally unstable (see Supplemental Material). 
The control operation then 
``pushes''   the spin configuration towards  ${\bf S}_j^{0}$ 
while maintaining the norm of $|{\bf S}_j(t)|=1$ at all times through
\begin{equation}
    {\bf S}_j(t') \rightarrow \frac{(1-a){\bf S}^{0}_j + a{\bf S}_j(t)}{| (1-a){\bf S}^{0}_j + a{\bf S}_j(t)|}~,
    \label{eq:control}
\end{equation}
with $0\leq a\leq 1$.  Thus, at small $a$ the control action is strong, while for $a$ near 1 it is weak.  

This control step is applied after each Hamiltonian time evolution step.  These two operations together make up one full time step.
In the following, we average over $N_{\mathrm{sample}}=1000$ 
different initial conditions, using chain lengths ranging from $L=32$ to $2000$, time-evolving for up to $L^2$ time steps.  The random initial states have each spin independently and uniformly distributed on the unit sphere.

We study the {\it activity}, i.e. the local deviation from the unstable fixed point ${\bf S}_i^0$ in Eq.~\eqref{eqn:spinspiral} defined as
\begin{equation}
    S_{\mathrm{diff}}(t,L;a)= \frac{1}{L}\sum_{i=1}^L\left [|{\bf S}_i(t)-{\bf S}_i^0| \right ]\equiv \frac{1}{L}\sum_{i=1}^L\left [ \delta \mathcal{S}_i(t) \right ],
    \label{eq:Sdiff}
\end{equation}
where we are denoting the average over initial states as $[\dots ]$. 
When $\delta \mathcal{S}_i$ is non-zero, the spin $i$ is locally away 
from the unstable fixed point that we are controlling onto. When $\delta \mathcal{S}_i =0$ for all $i$ the system is controlled, as shown in Fig.~\ref{fig:model}.

To ascertain the chaotic properties of the dynamics we evaluate the leading Lyapunov exponent $\lambda$.
To do so, we evaluate $\lambda$ following the approach of Ref.~\cite{benettin1976kolmogorov}; for this we create a copy of the system  with a very slightly different spin configuration.  We then evolve both systems, keeping track of the exponential growth or decay of the difference between the copy and the original system.  The spin configuration of the copy is regularly moved closer to or farther from that of the original system in order to keep the very small difference of a numerically convenient magnitude 
(see Supplemental Material).  

{\it Phases of the model}:
We now turn to the numerical solution of this model through analyzing the nature of its dynamics. In Fig.~\ref{fig:model} we show each phase of the model through space-time plots of the activity $\delta \mathcal{S}_i(t)$. For small values of $a$, the model is in the control phase, dominated by the control map. As a result, the activity vanishes at late times, with chaotic-like evolution only at short times. On the other hand, at large values of $a$ chaotic evolution takes over, with dynamics in the activity persisting at all points in space and time. 
For $a$ very near to but above the critical point $a_c$, the activity dynamics shows sustained directed percolation dynamics with spreading, splitting 
and local ``death'' of spatially sparse chaotic regions.  

We begin with considering $S_{\mathrm{diff}}(t)$ defined in Eq.~\eqref{eq:Sdiff} in the control phase, $a< a_c$. As shown in Fig.~\ref{fig:Sdiff_vs_t}(a), we find a time scale $t^*(L,N_{\mathrm{sample}})$ that separates two temporal regimes,
\begin{equation}
S_{\mathrm{diff}}(t,L;a) \sim
\begin{cases}
e^{-t/\xi_t}, & \xi_t < t \ll t^*, \\
e^{-t/\tau_C} & t \gg t^*,
\end{cases}
\label{eq:t*2}
\end{equation}
where $\tau_C$ is a property of the control map.
At intermediate times, we find that the decay follows the correlation time $\xi_{t}$, which diverges at $a_c$, and we  extract $\xi_t(a,L)$ through exponential fits, see dashed lines in Fig.~\ref{fig:Sdiff_vs_t}(a).  In this regime, $S_{\mathrm{diff}}$ is proportional to the fraction of samples that remain transiently chaotic, not having been yet ``captured'' by the control on to the fixed point.  The crossover time $t^*$ is when the final sample gets captured.  Thus  
at times past $t^*$ the decay of the activity is a simple exponential that is set by the Lyapunov exponent of the control map (being equal to $1/\tau_C=-\log(a)$, see dashed-dotted line in Fig.~\ref{fig:Sdiff_vs_t} (a), and the Supplemental Material.) 
The crossover time depends on system size $L$ and the number of samples $N_{\mathrm{sample}}$ as the long time decay of $S_{\mathrm{diff}}(t)$ is dominated by the sample(s) with the largest capture time(s). In the thermodynamic limit (i.e. $L\rightarrow \infty$ and $N_{\mathrm{samples}}\rightarrow \infty$) we  find that that 
$t^* \sim \log(LN_{\mathrm{samples}})$, consistent with the expectation that the probability distribution of capture times should follow a Poisson process. We have verified this in finite size simulations with a finite number of samples (see the Supplemental Material).

{\it Critical Properties}:
Near the transition,  
we study the activity  using the scaling hypothesis for dynamical critical phenomena~\cite{goldenfeld2018lectures}  
\begin{equation}
    S_{\mathrm{diff}}(t,L; a) \sim t^{-\delta}\, \Phi\left(\Delta\, t^{1/\nu_t},\, t/L^z\right),
    \label{eq:general_scaling}
\end{equation}
where $\Delta = (a_c - a)/a_c$ is the distance from the transition, $z$ is the dynamic exponent, $\nu_t$ is the  temporal correlation length critical exponent defined by $\xi_t\sim |\Delta|^{-\nu_t}$, $\Phi(x,y)$ is a universal scaling function, and 
$\delta \equiv \beta/\nu_t$. In the regime $t/L^z \ll 1$ the scaling is 
$
    S_{\mathrm{diff}}(t, L; a) \sim t^{-\delta}\, \Phi\left(\Delta\, t^{1/\nu_t},0\right)
    $
and at $\Delta=0$, following~\cite{mendonca2011monte} this implies  that
$    \log_b\left[
S_{\mathrm{diff}}(t/b)/S_{\mathrm{diff}}(t)
\right] \approx \delta,
$
We use this to independently determine $a_c = 0.7580(3)$ as well the critical exponents $\beta = 0.23(2)$ and $\nu_t = 2.2(2)$: Fig.~\ref{fig:Sdiff_vs_t}(b).  
For completeness we define $\nu \equiv \nu_t/z$ as the spatial correlation length critical exponent.

Sitting at the critical point, $\Delta=0$ we  perform scaling collapse using Eq.~(\ref{eq:general_scaling}) at $a_c=0.7580$ and get $z = 1.30(15)$. 
We now turn to how $\xi_t$ diverges as we approach $a_c$. Using the scaling hypothesis this leads to 
\begin{equation}
    \xi_t(a,L)\sim L^z f((a-a_c)L^{z/\nu_t})~,
    \label{eq:xi-scaling}
\end{equation}
where $f(x)$ is a universal scaling function.
As shown in Fig.~\ref{fig:Sdiff_vs_t}(c), we find $a_c = 0.7580$ and $\nu_t = 2.24$, which yields $z=1.34(10)$ in good agreement with our estimate from $S_{\mathrm{diff}}$.  We summarize the obtained exponents in Table~\ref{tab:exponents}.

\begin{figure}[t!]
   \centering
\includegraphics[clip, trim=40 45 46 45, width=0.43\textwidth]{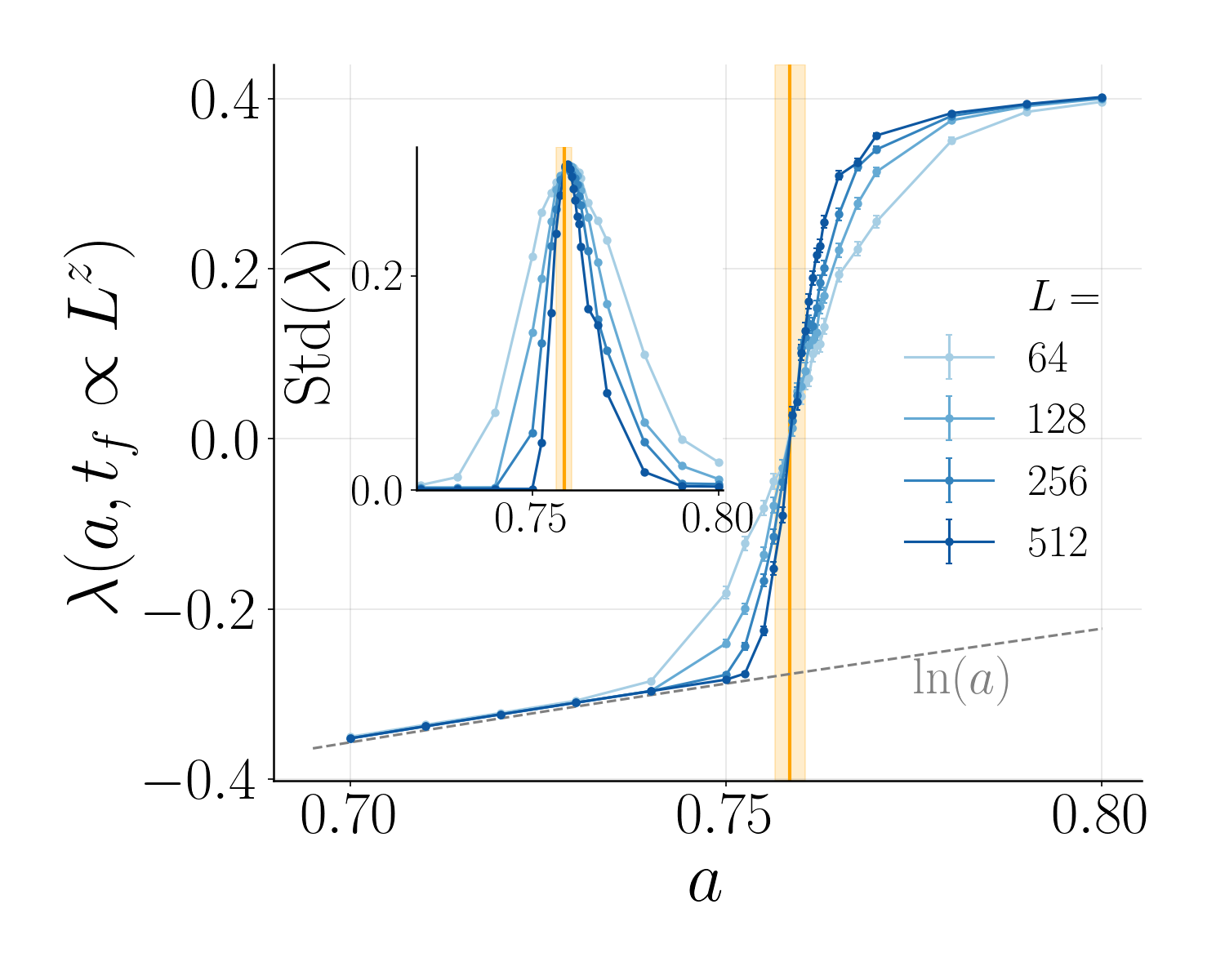}
\caption{{\bf Jump in the leading Lyapunov exponent and its fluctuations}: The average Lyapunov exponent $\lambda$ in Eq.~\eqref{eq:lambda} over 1000 samples evaluated at a late time $t_f \sim L^{1.34}$ by averaging over the time window $[t_f, 3t_f]$. As the critical point is approached from the chaotic phase $(a>a_c)$, $\lambda$ approaches a positive value $\lambda \approx 0.35$, while in the control phase $(a<a_c)$ the  $\lambda$ is close to $\ln(a)$ (gray dashed line), but averaging over the random signs in $J_{x,y}(t)$ from Eq.~\eqref{eq:ham} makes this value slightly larger. At the transition $a_c$ our data is consistent with a discontinuous jump in the Lyapunov exponent of an infinite system from its chaotic value to its control value, thus first-order-like behavior.  Note the crossing of the different $L$ curves that occurs roughly at $a = 0.7585 \pm 0.0010 \approx  a_c$ and $\lambda = 0$. The inset shows the standard deviation of $\lambda$ across many runs. Consistent with the jump in the Lyapunov exponent, the standard deviation displays a peak near $a_c$ with its maximum fixed over several $L$ and its width becoming narrower as $L$ increases.
}
\label{fig:lambda}
\end{figure}

Last, we come to the average and standard deviation of the leading Lyapunov exponent, denoted $\lambda$ and $\mathrm{std}(\lambda)$, respectively out to a late time $t_f=L^z\approx L^{1.34}$ and only average over times near $t_f$ from the range $[\frac{1}{2}t_f, 2t_f]$ (see Eq.~\eqref{eq:lambda} and Fig.~\ref{fig:lambda}).  We find that the average Lyapunov exponent jumps from a stable chaotic value close to $0.35$ for $a$ just above $a_c$ to a negative value in the control phase for $a<a_c$ that is slightly above $\ln(a)$ (due to averaging over random signs in $J_{x,y}(t)$ in Eq.~\eqref{eq:ham}). This first-order-like jump is distinct from the behavior of the activity, which goes to zero continuously, as shown in Fig.~\ref{fig:sdiff_vs_a} in the end matter. This produces a sharp peak in $\mathrm{std}(\lambda)$ with a $L$-independent maximum. These data collapse well onto universal scaling functions using the previously obtained critical exponents (Supplemental Material).

The temporal-randomness version of the Harris criterion \cite{PhysRevE.93.032143} tells us that such temporal randomness, as we have in our model, is relevant if $\nu_t<2$, as is the case for 1+1-dimensional directed percolation.  Thus we expect our model's critical point to be instead in a temporally-random universality class, which should have $\nu_t\geq 2$ by the temporal version of Ref. \cite{ccfs1}, as indeed we find.

\begin{figure}[t!]
    \centering
    \includegraphics[clip, trim=10 2 2 2, width=1\columnwidth]{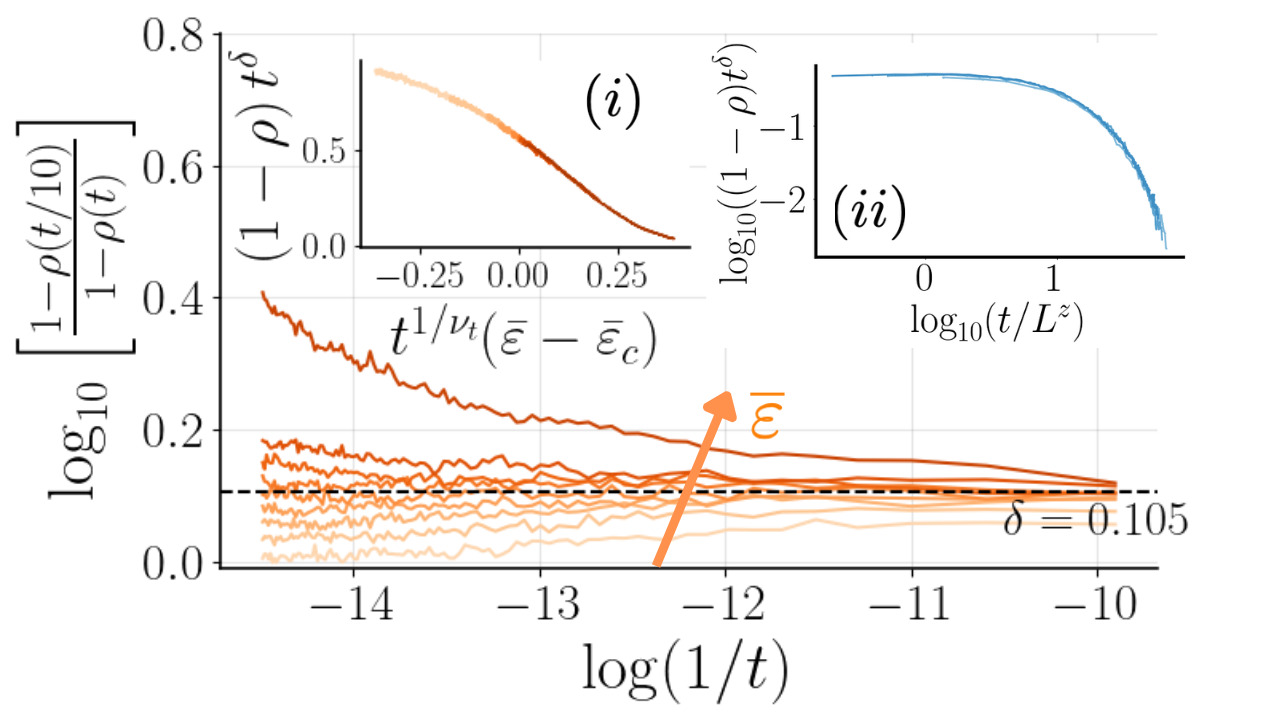}
    \caption{\textbf{Determination of critical exponents for Stavskaya's model with temporal randomness}: 
    (a) Using $L = 20000$ for $0.25315 \leq \overline{\varepsilon} \leq 0.25435$ to determine $\bar{\varepsilon}_c = 0.253725(5)$ and $\delta = 0.105(5)$;  see end matter for definition of the activity in this model ($\rho$). Left inset (i): Scaling collapse near $\bar{\varepsilon}_c = 0.253725$ to determine $\nu_t = 2.25(5)$, and therefore $\beta = 0.236(5)$.
    Right inset (ii): Finite-size scaling at $\bar{\varepsilon}_c = 0.253725$ with $L = 4000, \ 8000, \ 16000$ to determine $z = 1.45(5)$.}
    \label{fig:broad_stav}
\end{figure}

\begin{table}[t!]
    \centering
    \begin{tabular}{l|c|c|c}
        \hline\hline
        Model & $\beta$ & $\nu_t$ & $z$ \\
        \hline
        Spin chain  & $0.23(2)$ & $2.24(15)$ & $1.34(15)$ \\
        Time-Random Stavskaya  & $0.236(6)$ & $2.25(5)$ & $1.45(5)$\\
        Directed Percolation (DP) & $0.276$ & $1.734$ & $1.58$\\
        \hline\hline
    \end{tabular}
    \caption{Comparison of critical exponents across the two models with temporal randomness investigated in this work and the standard 1+1-dimensional directed percolation universality class. }
    \label{tab:exponents}
\end{table}

{\it Stavskaya's automaton:} The qualitative behavior of individual trajectories at our model's phase transition is reminiscent of directed percolation (DP): the system consists of spacetime regions that are locally either strongly chaotic or fully controlled. However, as detailed in Table~\ref{tab:exponents}, the critical exponents are distinct from standard directed percolation. Since a feature of our dynamics 
is the presence of temporal randomness, and the directed percolation transition is indeed unstable to temporal randomness (by the Harris criterion $\nu_t < 2$~\cite{ccfs1,PhysRevE.93.032143,Shkolnik-2025}), a natural conjecture is that adding temporal randomness to directed percolation drives it to the critical point that we have observed. To test this, we have explored the effects of temporal randomness on a standard model of 1+1-dimensional directed percolation, Stavskaya's automaton~\cite{stavskaya1968homogeneous,shnirman1968ergodicity,toom1968family}. The dynamics of this automaton is reviewed in the end matter. 
In short, each site has two states, which can be regarded as ``infected'' and ``healthy''.  The update consists of two steps, which are repeated: first each infected site deterministically infects its neighbor to the left, then each now-infected site becomes healthy with probability $\varepsilon$.  The configuration with only healthy sites is invariant, but below a certain $\varepsilon_c$ it is unstable to the introduction of infected sites. A typical random initial state then reaches a steady state that has a nonzero density of infected sites; this density is analogous to the ``activity'' we defined above. The transition at $\varepsilon_c$ is in the directed percolation universality class. We have explored a model in which the parameter $\varepsilon$ fluctuates in time (but is constant across the system at any time), considering the average value, $\overline{\varepsilon}$, to be the control parameter (see end matter for the details on randomness used). We analyzed the critical point of this temporally-random model using the same approach as for our spin chain model; the results are shown in Fig.~\ref{fig:broad_stav}: our estimates for critical exponents are consistent with those in the spin chain, see Table~\ref{tab:exponents}. Note that this new critical point satisfies the temporal CCFS bound as $z\nu=\nu_t\approx 2.25>2$, but does not saturate the bound. 

{\it Discussion:} We have introduced an absorbing-state transition with a mixed character: the mean activity vanishes continuously at the transition, and spatio-temporal correlations of the activity diverge with critical exponents consistent with temporally-random directed percolation---a universality class that has been little studied 
~\cite{IwanDPtemprand}. On the other hand, the leading Lyapunov exponent jumps discontinuously across the transition.  We conjecture, based on our results, that related control transitions without temporal randomness 
would exhibit a directed-percolation transition, and that the Lyapunov exponent would also jump discontinuously across this transition. 
This discontinuity of the Lyapunov exponent can be understood intuitively: in the control phase or even slightly in the chaotic phase, a configuration that is close \emph{everywhere} to the fixed point contracts to the fixed point at a finite rate (set by $\ln(a)$), so the Lyapunov exponent is negative. However, a typical chaotic configuration is \emph{far} from the fixed point on some sites, which seed chaos. As the transition is approached, chaotic regions become increasingly sparse in space, but remain locally chaotic, so the leading Lyapunov exponent remains positive and does not approach zero at the transition.  We expect that if one obtained the full Lyapunov spectrum, the {\it number density} of positive Lyapunov exponents should approach zero at the transition, while the largest exponent remains positive.  These qualitative features seem likely to persist in chaotic models without temporal randomness, and we conjecture that such models could exhibit a directed percolation transition at which the leading Lyapunov exponent also jumps. 

Our results suggest several avenues for future work: most obviously, exploring other nonequilibrium critical phenomena (e.g., Kardar-Parisi-Zhang universality~\cite{kpz1,kt_review}) in the presence of temporal randomness, and exploring the behavior of the Lyapunov exponent at other classical transitions out of chaos. An interesting family of such transitions are ``synchronization transitions''~\cite{ap2002}, between two sets of spins evolving under identical chaotic maps and also dissipatively coupled to each other. 
Whether the Lyapunov exponent jumps at these transitions or exhibits critical scaling is an interesting question for future work. Finally, the extension of our results to semiclassical systems~\cite{ap_review,MP2026,MP2026} with large but finite spins on every site~\cite{DL2014}, remains to be explored. 

\acknowledgments{{\it Acknowledgments}: We thank Sriram Ganeshan, Tom Iadecola, and Justin Wilson for useful discussions.  This work was partially supported by the US-ONR grant No.~N00014-23-1-2357 (J.H.P.),   the Rutgers Global International Collaborative Research Grant (J.H.P, M.K.). This work was performed in part at the Aspen Center for Physics, which is supported by National Science Foundation grant PHY-2210452 (J.H.P., S.G.) and at the Kavli Institute for Theoretical Physics (KITP), which is supported by grant NSF PHY-2309135 (J.H.P., S.G.). M.K. acknowledges support from the Department of Atomic Energy, Government of India, under project no. RTI4001.  J.H.P. acknowledges the hospitality of the International Centre for Theoretical Sciences (ICTS), Bangalore, India under the associateship program.



\bibliography{refs}
\clearpage

\section{\underline{End Matter}}
\setcounter{equation}{0}
\setcounter{figure}{0}
\renewcommand{\theequation}{EM\arabic{equation}}
\renewcommand{\thefigure}{EM\arabic{figure}}

\subsection{Activity across the Transition}

Using the estimate of $z=1.34$ we consider the activity out to final times $t_f=L^z$ as shown in Fig.~\ref{fig:Sdiff_vs_t}. In the chaotic phase, the activity saturates to a finite value in the thermodynamic limit at late times, where as in the control phase it goes to zero. The activity demonstrates a smooth transition from $0$ to positive finite value across the transition. Near the transition $a_c$, it follows the scaling ansatz.
\begin{equation}
    S_{\mathrm{diff}}(t,L,a)\sim \frac{1}{L^{\beta z/\nu_t}}g\left[(a-a_c)L^{z/\nu_t},t/L^z\right].
    \label{eqn:Sdiffscaling}
\end{equation}
Using the critical exponent values obtained in the main text, we preform this collapse [Fig.~\ref{fig:sdiff_vs_a} (inset)]. We find values for the critical exponents that agree well with what we obtain in the main text.

\begin{figure}[h!]
    \centering
    \includegraphics[clip, trim=43 35 22 31, width=1\linewidth]{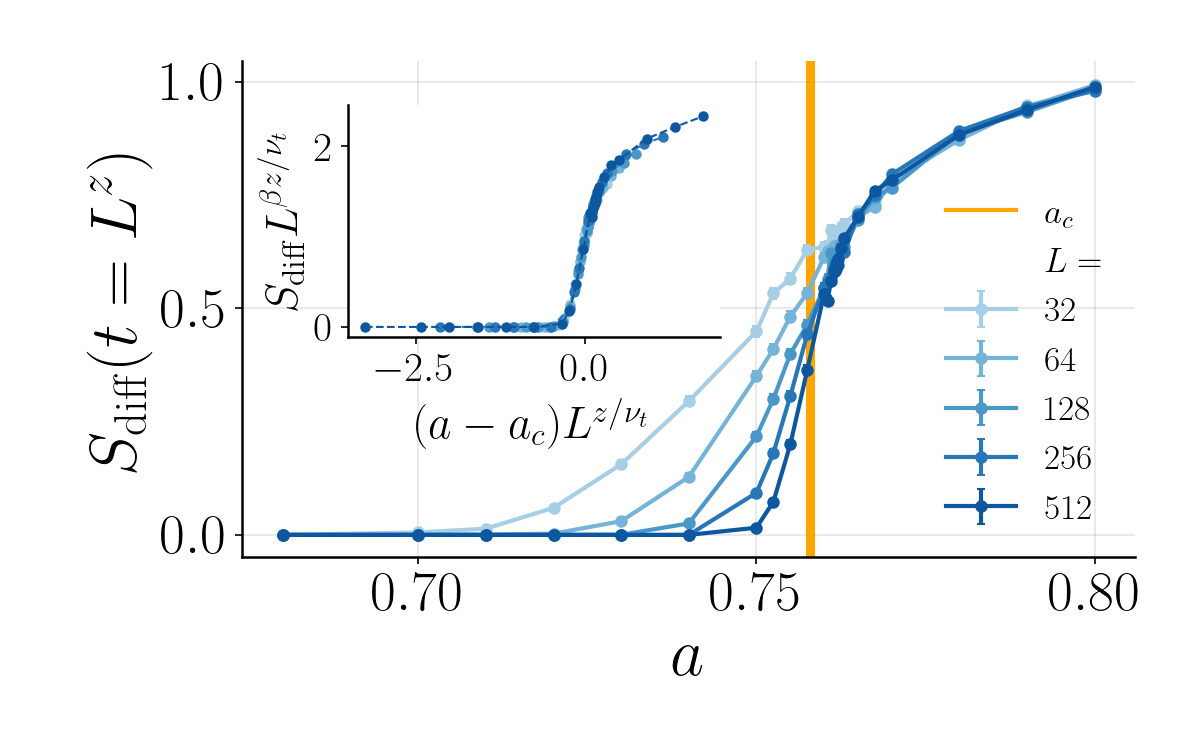}
    \caption{The activity at $t=L^{z=1.34}$ as a function of $a$. Note the smooth transition from $0$ to positive finite value across the transition. The inset shows FSS of the activity, using values obtained in Fig.~\ref{fig:Sdiff_vs_t}: $a_c = 0.758$, $\nu_t = 2.24$, $z = 1.34$, and $\beta = 0.23$.}
    \label{fig:sdiff_vs_a}
\end{figure}

\subsection{Analysis of Stavskaya’s model}

In this section of the end matter, we present an analysis of a conventional model for directed percolation that is subjected to temporal disorder. As directed percolation is unstable to temporal disorder, here we uncover the critical point that this flows to. As this agrees well with critical exponents found in the main text, our data implies, to our numerical precision, that these are in the same universality class.

To begin, we review Stavskaya's model, a well known model for directed percolation (DP). This  is a one-dimensional lattice of binary variables $\eta_i \in \{0, 1\}$~\cite{mendonca2011monte}. At every discrete time step, each site is updated according to the following:  
\begin{equation}
\eta_i(t+1) =
\begin{cases}
1, & \text{with probability } \varepsilon, \\
\eta_i(t)\,\eta_{i-1}(t), & \text{with probability } 1 - \varepsilon.
\end{cases}
\end{equation}
The model has a DP transition as $\varepsilon$ is tuned.
%
The order parameter for the transition is known to be the activity
\begin{equation}
    1-\rho(t,L;\varepsilon) = 1-\frac{1}{L}\sum_{i=1}^L\eta_i(t).
\end{equation}


There exists a critical $\varepsilon_c$ such that $1-\rho \to 0$ for $\varepsilon > \varepsilon_c$ but if $\varepsilon < \varepsilon_c$, then  $1-\rho$ goes to a finite value. By assuming the scaling ansatz near the critical point

\begin{equation}
    1-\rho(t,L; \varepsilon) \sim t^{-\beta/\nu_t} \Phi(\Delta_{\varepsilon} t^{1/\nu_t}, t/L^z) \,,
\end{equation}
we reproduce the work in \cite{mendonca2011monte} which yields critical exponents $\beta = 0.27(1)$, $\nu_t = 1.73(5)$, and $z = 1.60(5)$ showing that the phase transition belongs to the DP universality class (see Fig.~\ref{fig:stav_rep}).

We now introduce temporal disorder by replacing $\varepsilon$ with $\varepsilon(t) = 0.5(X(t))^n$, where $n$ is fixed at the beginning of a run to determine the average strength of the control, $X(t)$ chosen at each time step is uniformly distributed in $[0, 1]$. Now, we parameterize the strength of $\varepsilon(t)$ through its average value $\bar{\varepsilon} = a/(n+1)$.

Near the critical point, we use  the same scaling ansatz
\begin{equation}
    1-\rho(t,L; \bar{\varepsilon}) \sim t^{-\beta/\nu_t} \Phi(\Delta_{\bar{\varepsilon}} t^{1/\nu_t}, t/L^z) 
\end{equation}
where $\Delta_{\bar{\varepsilon}}=(\bar{\varepsilon}-\bar{\varepsilon}_c)/\bar{\varepsilon}_c
$ and we perform the same finite-size scaling procedure as we did in the main text on the spin chain to determine the critical exponents (see Fig.~\ref{fig:broad_stav}), the results of which are shown in Table~\ref{tab:exponents}. We see good agreement between these two vastly different models, providing good evidence for the hypothesis that our spin chain transition belongs to a temporally random DP universality class.

\begin{figure}[t!]
    \centering
    \includegraphics[clip, trim=6 25 30 30, width=0.9\columnwidth]{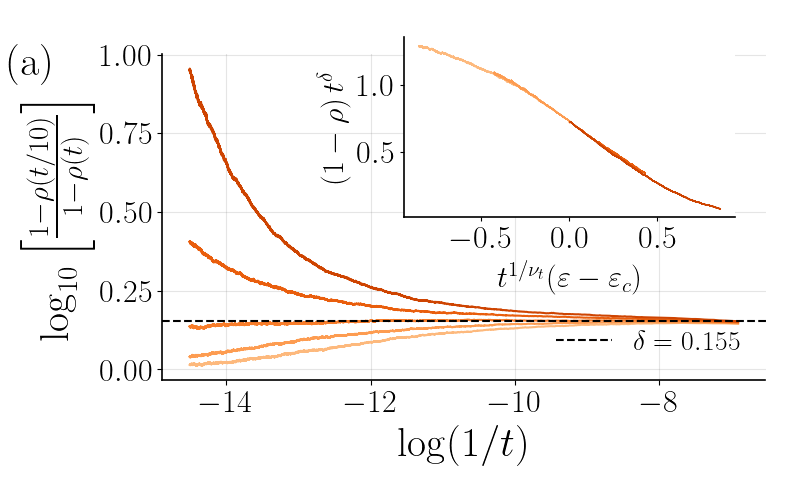}
    \includegraphics[clip, trim=22 22 30 30, width=0.9\columnwidth]{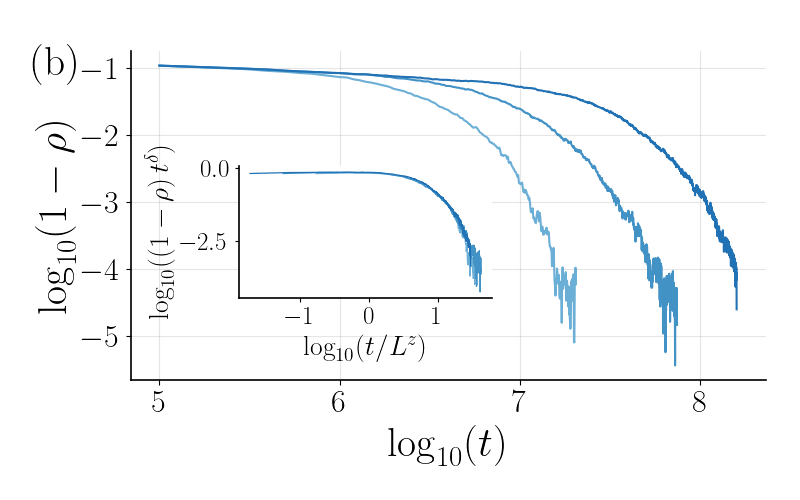}
    \caption{\textbf{Determination of critical exponents for Stavskaya's model} 
    (a) Using $L = 20000$ at $\varepsilon \in \{0.2943, 0.2944, 0.2945, 0.2946, 0.2947\}$ to determine $\varepsilon_c = 0.29450(5)$ and $\delta = 0.155(5)$. Inset shows scaling collapse near $\varepsilon_c = 0.29450$ to determine $\nu_t = 1.735(5)$, and therefore $\beta = 0.27(1)$.
    (b) Finite-size scaling with $L = 4000, \ 8000, \ 16000$ at $\varepsilon_c = 0.29451$ to determine $z = 1.60(5)$.
    }
    \label{fig:stav_rep}
\end{figure}

\clearpage
\onecolumngrid
\section{\underline{Supplemental Material}\\ ``Control transition in a temporally random classical spin chain"}

\setcounter{equation}{0}
\setcounter{figure}{0}
\renewcommand{\theequation}{S\arabic{equation}}
\renewcommand{\thefigure}{S\arabic{figure}}

\section{Computing the leading Lyapunov Exponent}
In this section, we discuss the computation of the leading Lyapunov exponent. 
We discretize our evolution following Ref.~\cite{benettin1976kolmogorov} and define the leading Lyapunov exponent as
\begin{equation}
\lambda = \frac{1}{n\tau} \sum_{j=1}^{n}\ln\bigg(\frac{|d_j|}{\epsilon} \bigg). 
\label{eq:lambda}
\end{equation}
We take $\epsilon=0.01$ and $\tau$ is a discretization with time $t_j=nj$ (here we take $\tau=1/J$). 

The final time is given by $t_f=n \tau$ where $n$ is the number of steps. Initial times that are affected by the initial state are dropped from the sum to obtain a more accurate estimate of $\lambda$. We initialize our system of interest, call it system $A$, in a random initial state and we then construct a copy of the system $B$ that is very close to system $A$, where ${\bf S}_i^B={\bf S}_i^A$ for all $i$ expect $i=L/2$ (i.e., the middle site). Here, we randomize each direction of the middle site within the window of $\epsilon^{\alpha}\in [-\epsilon,\epsilon]$ defining the vector $\bm{\epsilon}$ such that ${\bf S}^B_{L/2}=({\bf S}_{L/2}^A+\bm{\epsilon})/|({\bf S}_{L/2}^A+\bm{\epsilon})|$. After a short period of time, such that the deviation between spins in copy $A$ and $B$ have spread across the chain, we evaluate the quantity 
\begin{equation}
    |d_j|=\sqrt{\sum_{i=1}^L({\bf S}^A_i(t_j)-{\bf S}^B_i(t_j))\cdot ({\bf S}^A_i(t_j)-{\bf S}^B_i(t_j))}.
\end{equation}
Last,  we  reset copy $B$ to be within $\epsilon/L$ of copy $A$ via~\cite{benettin1976kolmogorov}
\begin{equation}
\label{eq:sbsa}
{\bf S}_i^{B,\rm{reset}} = \frac{{\bf S}_i^A + \frac{\epsilon}{|d_i|} ({\bf S}_i^B-{\bf S}_i^A)}{|{\bf S}_i^A + \frac{\epsilon}{|d_i|} ({\bf S}_i^B-{\bf S}_i^A)|}.
\end{equation}

Another estimate of $z$ can be obtained by analyzing the standard deviation of $\lambda(t)$ across many runs (see Fig.~\ref{fig:fss_std_lambda_t}).
$
    \mathrm{Std}\left(\lambda(t, L, a=a_c)\right),
$
where 
$
    \lambda(t) \equiv \frac{1}{\tau}\ln(\frac{|d_{j=t}|}{\epsilon}).
$

\begin{figure}[h]
    \centering
    \includegraphics[clip, trim=30 30 30 33, width=0.5\linewidth]{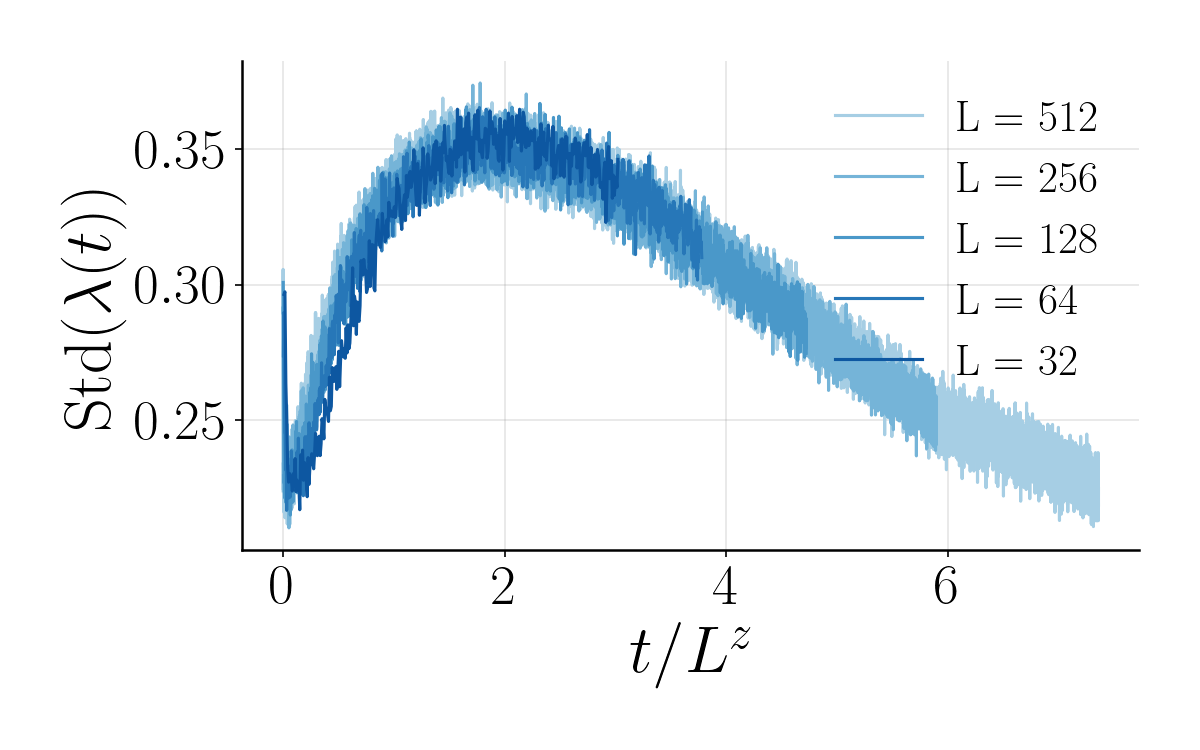}
    \caption{FSS of $\mathrm{Std}(\lambda(t))$ at $a_c = 0.7588$ which gives $z \approx 1.38 \pm 0.1$.
    }
    \label{fig:fss_std_lambda_t}
\end{figure}

\begin{figure}[h]
    \centering
    \includegraphics[trim={10 20 20 10}, clip, scale=0.4]{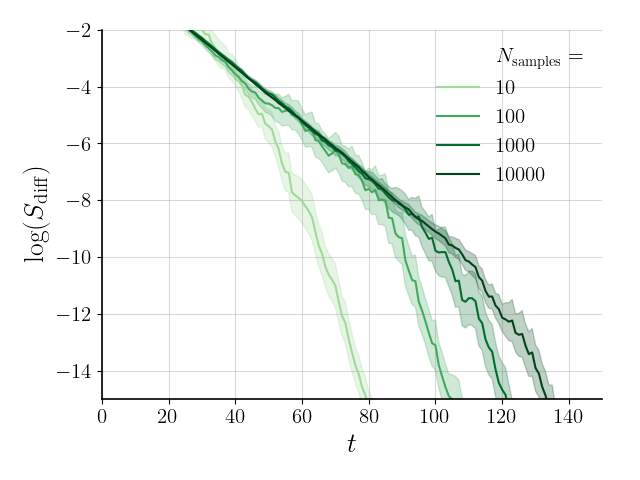}
    \caption{$\log(S_{\mathrm{diff}}(t))$ as a function of time for fixed $L = 256$ and increasing $N_{\mathrm{samples}}$. The samples for $N_{\mathrm{samples}} = 10, 100$ were picked by hand as outlier long lived samples can dominate and give the impression that $t^*$ occurs at a later time. The data shows clearly that $t^*$ increases when $N_{\mathrm{samples}}$ increases.}
    \label{fig:cherried_s_diff_vs_IC}
\end{figure}

\begin{figure*}
\includegraphics[trim={10 0 20 10}, clip, scale=0.25]{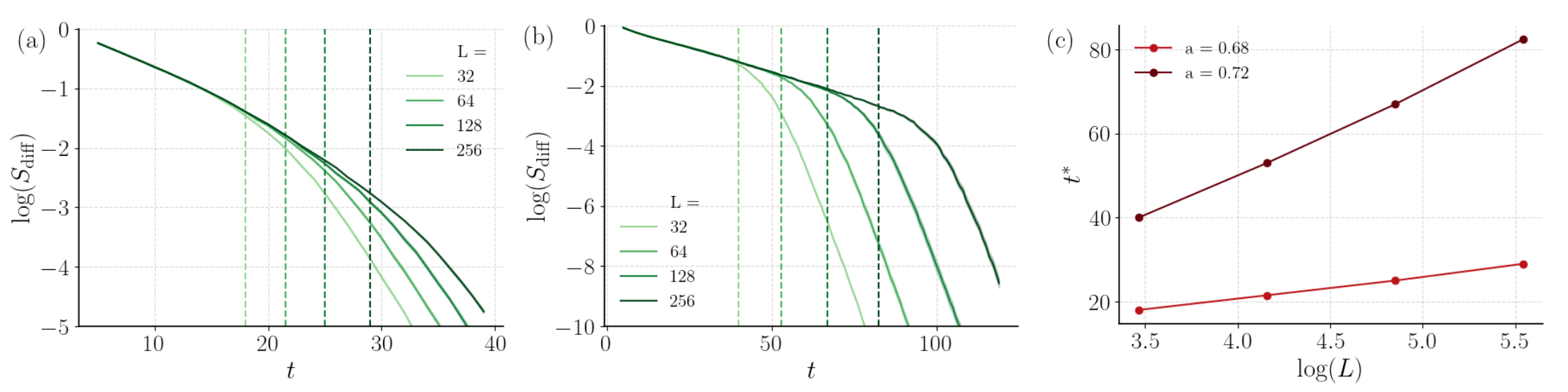}
\caption{(a) $\log(S_{\mathrm{diff}}(t,a=0.68))$ for different $L$ values showing a clear correlation between $t^*$ and $L$. (b) Same as (a), but for $a = 0.72$. (c) $t^*$ as a function of $\log(L)$ showing that the relationship between $t^*$ and $\log(L)$ is indeed linear when $N_{samples}$ is held fixed, as per Eq.~\eqref{eq:t*_vs_N_and_L}.}
\label{fig:log_s_diff_vs_L}
\end{figure*}

\section{Estimate of $t^*$}

As stated in the main text, we expect 

\begin{equation}
    t^*\sim \log(LN_{\mathrm{samples}}) = \log(L) + \log (N_{\mathrm{samples}})
    \label{eq:t*_vs_N_and_L}
\end{equation} 
We verify this by taking $L$ constant and allowing the number of samples to vary (Fig.~\ref{fig:cherried_s_diff_vs_IC}) and by holding $N_{\mathrm{samples}}$ constant while allowing 
$L$ to vary (Fig.~\ref{fig:log_s_diff_vs_L}). Fig.~\ref{fig:cherried_s_diff_vs_IC} only to shows that the relationship holds qualitatively, while Fig.~\ref{fig:log_s_diff_vs_L} goes further and shows a clear linear relationship between $t^*$ and $\log(L)$.

\section{Control Phase Dynamics}

\begin{figure}[t!]
    \centering
    \includegraphics[trim={10 20 20 10}, clip, scale=0.391]{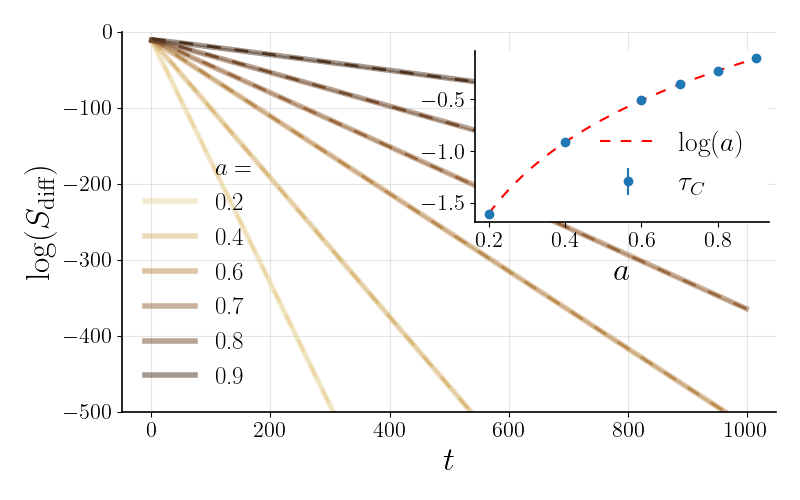}
    \caption{Showing an average over 1000 initial conditions of $\log(S_{\mathrm{diff}})$ as a function of time when we start the initial conditions at a fixed $S_{\mathrm{diff}}(t=0) = \epsilon = 10^{-5}$ and $L = 128$. The main figure shows linear fits of the data and the inset plots the slopes of the fits as a function of the corresponding $a$ value against $\log(a)$ shown as a red dashed line (not a fit), which agrees well with the data.}
    \label{fig:control_epsilon_exp}
\end{figure}

In order to separately estimate $\tau_C$, instead of taking random initial conditions we take a random initial spin configuration that is $\epsilon$ away from the unstable fixed point, in particular we take
${\bf S}_j^{\mathrm{init}}=({\bf S}_j^0+\epsilon {\bf S}_j^r)/|({\bf S}_j^0+\epsilon {\bf S}_j^r)|$ 
where  
$\epsilon \ll 1$ and ${\bf S}_j^r$ is randomly distributed on the unit sphere. In the following we take $\epsilon = 10^{-5}$ and the results for $0<a<1$ are shown in Fig.~\ref{fig:control_epsilon_exp}. When restricting the initial conditions to an $\epsilon$ away from the unstable fixed point the model does not become chaotic and is always controlled. Here, we always work in the limit $\epsilon \rightarrow 0$, such that as we take $t\rightarrow \infty$ and $L\rightarrow \infty$ we can always take $\epsilon$ sufficiently small. From this we are able to estimate the decay into the control state directly. We  find excellent agreement for $\tau_C$ when the model is  deep in the control phase between fully random initial conditions   and those that are $\epsilon$ away from the unstable fixed point as shown in Fig.~\ref{fig:control_epsilon_exp} allowing us to conclude that
\begin{equation}
    1/t_C=\log(1/a). 
\end{equation}
This shows that the action of the control map when close to the unstable fixed point is to push it exponentially closer, larger $a$ pushes it faster. Importantly, $t_C$ is smooth through the transition once we take limited initial conditions near the unstable fixed point. Lastly, we also plot the cross over time scale $t^*$, which marks the cross over between the early and late times exponential decay (Fig.\ref{fig:log_s_diff_vs_L}.c).

\section{Removing Solitons}
\begin{figure}
   \centering
\includegraphics[trim={5 0 20 10}, clip,scale=0.22]{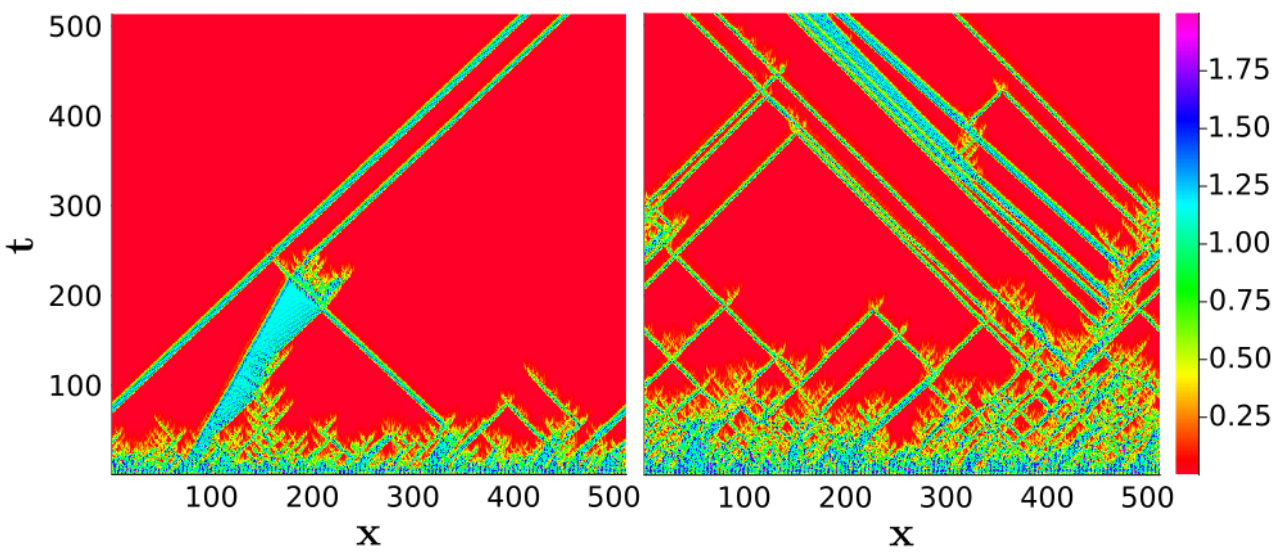}
\caption{
Activity dynamics revealed through space-time plots of $\delta \mathcal{S}_i(t)$ using the spin chain model in Eq.~\eqref{eq:EOM}. As we approach the chaotic phase we find ``soliton like'' excitations (shown as a non-dispersing line in (a)) within patterns of activity that look like  directed percolation. These soliton's proliferate as we exit the control phase and could introduce an intermediate phase between chaos and control. (a) is for $a = 0.5675$ and (b) is for $a = 0.6275$. The fact that these disturbances can exist for such small $a$ makes it difficult to define a control and chaotic regime.
}
\label{fig:solitons}
\end{figure}
When we run the dynamics with time independent $J_x(t)=J_y(t)=J$ we find effective solitons (i.e. moving with velocity $v_{\mathrm{soliton}}\approx \pm 1$ with a random sign without changing its shape) near the transition out of the control phase as shown the activity dynamics of $\mathcal{S}_x(t)$ in Fig.~\ref{fig:solitons}.
These solitons modify the phase diagram and potentially place an intervening phase between chaos and control, which is a further detail beyond our current intentions of studying a many-body chaos to control phase transition. Therefore, in the following we remove these solitons by making the dynamics more chaotic by making the sign of $J_x(t)$ and $J_y(t)$ in Eqs.~\eqref{eq:ham} and \eqref{eq:EOM} random at each application of the Hamiltonian dynamics. This is a simple way to remove the soliton like features completely as shown in Fig.~\ref{fig:model}  without strongly changing any of the energy scales of the problem. In the following all of the data shown is for  the dynamical model with random signs for $J_x(t)$ and $J_y(t)$, which is averaged over in each sample.

\section{Linear instability analysis around the $\pi/4$-spiral state}
Here we present a simple analytic theory that demonstrates the modularity of this system near the $N=4$ unstable spiral fixed point. We use the parameterization of Ref.~\cite{das2020} slightly modified
\begin{equation}
    S_j^x = s_j, \,\, S_j^y=f(s_j)\cos(\phi_j), \,\,
    S_j^z=f(s_j)\sin(\phi_j)
\end{equation}
where $f(x)=\sqrt{1-x^2}$. In these variables the equations of motion are derived from the Hamilton equations of motion $-d\phi_j/dt=\partial H/\partial s_j$ and $d s_j/dt = \partial H/\partial \phi_j$  (using $J\rightarrow -J$).
For uniform exchange $J=J_x=J_y=J_z$ we obtain
\begin{equation}
    \begin{aligned}
    \frac{d s_j}{dt}=Jf(s_j)f(s_{j+1})\sin(\phi_{j+1}-\phi_j) 
-Jf(s_j)f(s_{j-1})\sin(\phi_{j}-\phi_{j-1})
\end{aligned}
\label{eq:expanded_sj}
\end{equation}

and

\begin{equation}
    \begin{aligned}
         \frac{d \phi_j}{dt}=-J\bigg[f'(s_j)f(s_{j+1})\cos(\phi_{j+1}-\phi_j)
     +f'(s_j)f(s_{j-1})\cos(\phi_{j}-\phi_{j-1})\bigg] 
         -J(s_{j-1}+s_{j+1}).
    \end{aligned}
    \label{eq:expanded_phij}
\end{equation}

If we expand the parameters about our classical $N=4$ spiral fixed point

\begin{align}
    s_j^0 = 0\,, \quad \quad
    \phi_j^0 &= \frac{\pi}{2}j
\end{align}
such that $s_j = s_j^0+\delta s_j$, $\phi_j=\phi_j^0+\delta \phi_j$, Eq.~\eqref{eq:expanded_sj} and Eq.~\eqref{eq:expanded_phij} become

\begin{eqnarray}
    \frac{d \delta s_j}{d t} = 0\,, \quad
    \frac{d \delta \phi_j}{d t} &=& -J_x(\delta s_{i-1}+\delta s_{i+1}).
\end{eqnarray}

This has the in time solution 

\begin{equation}
    \phi_j(t)=-(\delta s_{j-1}+\delta s_{j+1})\,t+\pi j/2,
\end{equation}
with $\delta s_j(t)=\delta s_j(0)$ is fixed by the initial condition. Therefore, we see that the phase perturbations $\delta \phi_j$ grow linearly with time and the spin evolves periodically in time, not exponentially, hence this perturbation around the fixed point is marginal for the period 4 spiral.

\end{document}